\documentclass[runningheads]{svmult}

\usepackage{makeidx}   
\usepackage{graphicx}  
\usepackage{subeqnar}  
\usepackage{multicol}  
\usepackage{physprbb}  
\makeindex             

\begin{document}

\title*{Observations and Simulations of Relativistic Jets}

\toctitle{Observations and Simulations of Relativistic Jets}

\titlerunning{Observations and Simulations of Relativistic Jets}

\author{Jos\'e-Luis G\'omez}

\authorrunning{Jos\'e-Luis G\'omez}

\institute{Instituto de Astrof\'{\i}sica de Andaluc\'{\i}a (CSIC),
	Apartado 3004, Granada 18008, Spain}

\maketitle

\begin{abstract}
The recent improvement in VLBI arrays is providing information of the emission
and magnetic field structure of relativistic jets, both extragalactic and
galactic (microquasars), with unprecedented spatial and temporal
resolution. These observations are revealing the importance of the
hydrodynamical processes that govern the jet evolution, which can be studied
by the recently developed time--dependent relativistic hydrodynamical
models. Computation of the non--thermal emission from these hydrodynamical
models, and its comparison with actual sources, is proving as one of the
most powerful tools in the understanding of the physical processes taking
place in these jets. This paper reviews some of the recent observational
results, as well as the numerical models used to interpret them.

\end{abstract}

\section{Introduction}

Since the classical works of \cite{BR74} and \cite{BK79}, our knowledge of the
jet physics in AGNs and microquasars have improved significantly by analytical
and numerical models. The analytical efforts provided the basic frame work to
understand the non--thermal synchrotron and inverse Compton emission of
inhomogeneous jets (\cite{Al80}, \cite{Ko81}); spectral evolution of shock
waves, associated with the superluminal components (\cite{BK76}, \cite{MG85},
\cite{HAA85}); and polarization (e.g., \cite{CC90}). The implementation of
these analytical results into numerical models have allowed testing of the
basic jet model hypotheses, as well as a more detailed comparison with the
observations (\cite{Jo88}, \cite{HAA89}, \cite{HAA91}, \cite{JL93},
\cite{JL94a}, \cite{JL94b}). However, these early numerical models are limited
by the lack of a detailed non--linear model to study the relativistic jet
dynamics, being forced to adopt simplified stationary relativistic
hydrodynamical models.

On the other hand, Newtonian hydrodynamical numerical models have been used to
obtain a more detailed study of the jet dynamics, and its influence in the jet
observational properties. These models explored with great success the
morphology, dynamics and stability of jets (see e.g. reviews \cite{N96},
\cite{Fe98}), mainly aimed to study the large scale jet structure. However,
these models cannot account for the relativistic effects that are of special
importance in the overall emission of jets in AGNs and microquasars.

First studies of relativistic (magneto)hydrodynamical jets were obtained for
stationary flows (\cite{Wi87}, \cite{DP93}, \cite{Bo94}). A significant step
forward in the field of numerical simulations came with the development of
modern high--resolution techniques in numerical hydrodynamics, mading feasible
the computation of time--dependent simulations of relativistic jets
(\cite{vP93}, \cite{Ch94}, \cite{Ch95}, \cite{Ch97b}, \cite{DH94},
\cite{KF97}, and reviews \cite{Ch97a}, \cite{Ch99}, \cite{Ch00},
\cite{MM01}). These models are capable, for the first time, to study the jet
dynamics with unprecedented detail, and under very similar conditions as it is
thought are taking place in real sources (strong shocks, relativistic internal
energies and bulk flow velocities, etc.). Some of the latest simulations have
started to explore three dimensional relativistic jets (\cite{MA99},
\cite{MA99b}, \cite{MA00}, \cite{Ha00}), magnetized relativistic jets
(\cite{K99}), as well as jet formation and collimation making use of the first
general relativity magnetohydrodynamical codes (\cite{KSK98}, \cite{KSK99},
\cite{Koi00}, \cite{Me01}, \cite{MKU01}).

However, The emission structure that we observe in our VLBI images is not just
a direct mapping of the jet hydrodynamical variables (pressure, density,
velocity). The final radiation reaching our detectors is greatly determined by
other several processes, like Faraday rotation, opacity, particle
acceleration, radiative losses, and, most importantly, by relativistic effects
such as light aberration and light travel time delays. For relativistic speeds
(and small viewing angles) time delays can be of such importance as to render
the emission images with no apparent relationship to the hydrodynamical jet
structure. Therefore, the state of the art in the simulation of relativistic
jets involves the computation of the emission, taking into account the
appropriate relativistic and transfer of radiation processes, from the
relativistic time-dependent hydrodynamical results (\cite{JL95}, \cite{JL96a},
\cite{JL96b}, \cite{JL97}, \cite{HDM96}, \cite{KF96}, \cite{MHD97},
\cite{KF97}, \cite{MA99b}, \cite{MA00}, \cite{JRE99}, \cite{IJL01}, and review
\cite{JL01}). Comparison of these simulations with actual observations 
should provide a better understanding of the relativistic jets in AGNs and
microquasars.

\section{Relativistic HD and Emission models}

Most of the energy transported in relativistic jets is assumed to be carried
out by a population of thermal electrons. This population determines the
hydrodynamical evolution of the jet, and can be simulated by the relativistic
HD codes. However, the non-thermal emission observed from these jets is
originated by a second population of high energy, non-thermal
particles. Detection of circular polarization in the jet of 3C~279
(\cite{Wj98}), as well as in 3C~84, PKS~0528+134, and 3C~273 (\cite{Ho99}),
suggests that this non-thermal population is mainly composed by pairs
electron--positron. It is still unclear how this non-thermal population is
originated (\cite{Al96a}), perhaps by pair cascades (\cite{Ma93},
\cite{BL95}), neutron decay (\cite{EW78}, \cite{GK90}), or by acceleration of
the thermal electrons at a strong recollimation shock presumably associated
with the VLBI core (\cite{DM88}, \cite{Al98}, \cite{Mat98},
\cite{Mat00}). This population of non-thermal electrons is subsequently
re-accelerated at shocks along the jet (\cite{Ki98}, \cite{GAK99},
\cite{Ki00}, \cite{Ga01}), and incremented with contributions from thermal
electrons accelerated at the same shocks.

In order to compute the expected emission from the hydrodynamical models is
necessary to establish a relationship between the thermal and non-thermal jet
populations. A common assumption considers that the particle and energy density
of the non-thermal electrons is a constant fraction of the thermal
electrons (\cite{Ra77}, \cite{WS83}, \cite{JL95}, \cite{JL97}, \cite{MA00},
\cite{MHD97}, \cite{KF97}). The population of non-thermal electrons is 
assumed to share the same dynamics as the thermal population, which can
therefore be computed using the hydrodynamical simulations. Any exchange
between internal and kinetic energy along the jet will maintain the
proportionality between thermal and non-thermal populations. Only
non-adiabatic processes, such as gains by particle acceleration in shocks or
losses by radiation can modify this proportionality.

Radiative losses at radio wavelengths are expected to be small, except at
strong shocks, such as the terminal hot spots and jet cocoon. It is therefore
expected that computation of parsec scale radio emission will not be severely
influenced by changes in the non-thermal population produced by radiative
losses or particle accelerations. At higher energies (i.e., optical) and at
sites of strong shocks it is possible to trace the electron non-thermal
population gains and losses of energy by computing the electron energy
transport during the jet evolution. This has been recently considered for
non-relativistic magnetohydrodynamic simulations (\cite{JRE99}), allowing the
exploration of the effects induced in the emission by synchrotron aging and
electron energy gains at strong shocks.

To compute the synchrotron emission is necessary to distribute the internal
energy calculated from the hydrodynamic codes among the relativistic
non-thermal electrons. This is done by assuming a power low energy
distribution in the form $ N(E) $d$ E=N_{o}E^{-\gamma} $d$ E $, with $ E_{min}\leq
E\leq E_{max} $, and spectral index $\gamma$. Neglecting radiative energy losses
and particle accelerations, the ratio $C_{E}$ between the maximum and
minimum energy remains constant trough the computational domain and can be
considered a free parameter of the model. The power law is then fully
determined by the equation (\cite{JL95})
\begin{equation}
N_{o}=\left[ {{U\, (\gamma-2)}\over{1-C_{E}^{2-\gamma}}}\right]
^{\gamma-1}\left[ {{1-C_{E}^{1-\gamma}}\over{N\, (\gamma-1)}}\right] ^{\gamma-2}
\end{equation}
 and
\begin{equation}
E_{min}={{U}\over{N}}{{\gamma-2}\over{\gamma-1}}
{{1-C_{E}^{1-\gamma}}\over{1-C_{E}^{2-\gamma}}}
\end{equation}
where $ U $ and $ N $ represent the electron energy density and number
density, respectively, as calculated by the hydrodynamical codes.

It is still largely unknown what may be the role played by the magnetic field
in the jet dynamics of AGNs and microquasars. There are some evidence pointing
towards a small contribution of the magnetic field in the dynamics
(\cite{Hu01}), although only future observations and magnetohydrodynamical
simulations (\cite{DP93}, \cite{vP96}, \cite{K99}, \cite{KSK99}) could answer
this question. So far, and mainly due to the fact that the emission
computations have been performed for purely hydrodynamical models, the
magnetic field has been assumed to be dynamical negligible, with a magnetic
energy density proportional to the particle energy density (\cite{WS83}),
leading to a field with magnitude proportional to $\sqrt{U}$. Once the
magnetic field is considered dynamically negligible, ad-hoc magnetic field
structures can be considered. To account for the small degree of linear
polarization observed in many sources, the magnetic field is commonly
considered to be predominantly turbulent.

\subsection{Synchrotron Radiation Transfer}

The transfer of synchrotron radiation have been considered in detail
previously under different astrophysical scenarios, see e.g., \cite{Pac},
\cite{JO7a}, \cite{JO7b}, \cite{HAA89}. Its implementation for computing the
polarized emission from the hydrodynamical models can be summarized as follows
(\cite{JL93}, \cite{JL94a}, \cite{JL94b}, \cite{JL95}).

To obtain the emission and absorption coefficients for the transfer of
polarized synchrotron radiation let consider the direction of the component of
the magnetic field in the plane of the sky at a given computational cell be
specified as direction $2$, and let the axis $1$, $2$, and the direction
toward the observer be directions which form a right-handed orthogonal system
in that order. In this system, the emission and absorption coefficients,
respectively, are then computed {\it in the fluid frame} using (see e.g.,
\cite{Pac})

\begin{subeqnarray}
\varepsilon ^{(i )}_{\nu }={{\sqrt{3}}\over{16\pi
}}{{e^{3}}\over{mc^{2}}}\, C_{1}^{(\gamma-1)/2}\,N_{o}\,(B\sin
\vartheta)^{(\gamma+1)/2}\,\nu^{(1-\gamma)/2} \nonumber \\
\int ^{x_{max}}_{x_{min}} x^{(\gamma-3)/2}\,[F(x)\pm G(x)] \,{\rm d}x
\setcounter{eqsubcnt}{0}
\label{ecf}
\end{subeqnarray}

\begin{subeqnarray}
\kappa _{\nu }^{(i)}={{\sqrt{3}e^{3}}\over{16\pi
m}}\,(\gamma+2)\,C_{1}^{\gamma/2}\,N_{o}\,(B\sin \vartheta )^{(\gamma+2)/2} \, 
\nu^{-(\gamma+4)/2} \nonumber \\
\int ^{x_{max}}_{x_{min}} x^{(\gamma-2)/2}\,[F(x)\pm G(x)]\,{\rm d}x
\setcounter{eqsubcnt}{0}
\label{acf}
\end{subeqnarray}
where the plus sign is to be taken for $i$=1, and the minus sign is valid for
$i$=2; $ \vartheta $ is the angle between the magnetic field and the line of
sight; and
$$
C_{1}={{3e}\over{4\pi m^{3}c^{5}}}$$
$$
x={{\nu }\over{C_{1}B\sin \vartheta E^{2}}}$$
$$
F(x)=x\int ^{\infty }_{x}K_{5/3}(z)dz$$
$$
G(x)=xK_{2/3}(x)$$ where $ K_{5/3} $and $ K_{2/3} $are the corresponding
Bessel functions. 

If the distribution of the magnetic field within the source is not uniform in
orientation the $(1,2)$ system will differ from cell to cell, thus it is more
convenient to formulate the transfer equations in a system $(a,b)$, which is
fixed in orientation with respect to the observer. The relative orientation of
the axis $2$ with respect to the axis $a$, which defines the angle $\chi_B$,
will change from cell to cell as the magnetic field does.

The radiation field is characterized by the four Stokes parameters $I$, $Q$,
$U$, and $V$, or equivalently by $I^{(a)}$, $I^{(b)}$, $U$, and $V$, where
$I=I^{(a)}+I^{(b)}$ and $Q=I^{(a)}-I^{(b)}$. Provided jets in blazars exhibit
very low circular polarization we can assume $V=0$. $I$ is the total
intensity, and $Q$ and $U$ determine the degree of polarization $$\Pi =\left(
Q^{2} + U^{2}\right)^{1/2}$$ and the polarization position angle $$\chi =
\frac{1}{2} \; \arctan \left(\frac{U}{Q}\right).$$

The change of the parameters $I^{(a)}$, $I^{(b)}$ and $U$ characterizing the
radiation passing through a volume element of length ${\rm d}s$ can be
obtained by solving the transfer equations in the $(1,2)$ system and
transforming to the $(a,b)$ system, given by 

\begin{subeqnarray}
\frac{{\rm d}I^{(a)}}{{\rm d}s} = 
     I^{(a)} \left[ - \kappa^{(1)}_{\nu} \sin^{4}\chi_{B} - 
      \kappa^{(2)}_{\nu} \cos^{4}\chi_{B} - \frac{1}{2} \kappa_{\nu} \sin^{2}2\chi_{B} 
      \right] \nonumber \\
    + U \left[ \frac{1}{4} (\kappa^{(1)}_{\nu} - \kappa^{(2)}_{\nu}) \sin2\chi_{B} + 
     {\rm d}\chi_{{\rm F}}/{\rm d}s \right] \nonumber \\
    + \varepsilon^{(1)}_{\nu} \sin^{2}\chi_{B}+\varepsilon^{(2)}_{\nu} \cos^{2}\chi_{B}
\setcounter{eqsubcnt}{0}
\label{ia}
\end{subeqnarray}
\begin{subeqnarray}
\frac{{\rm d}I^{(b)}}{{\rm d}s} = 
      I^{(b)} \left[ - \kappa^{(1)}_{\nu} \cos^{4}\chi_{B} - 
      \kappa^{(2)}_{\nu} \sin^{4}\chi_{B} - \frac{1}{2} \kappa_{\nu} \sin^{2}2\chi_{B} 
      \right] \nonumber \\
    + U \left[ \frac{1}{4} (\kappa^{(1)}_{\nu} - \kappa^{(2)}_{\nu}) \sin2\chi_{B} -
      {\rm d}\chi_{{\rm F}}/{\rm d}s \right] \nonumber \\
    +\varepsilon^{(1)}_{\nu} \cos^{2}\chi_{B}+\varepsilon^{(2)}_{\nu} \sin^{2}\chi_{B}
\setcounter{eqsubcnt}{0}
\end{subeqnarray}
\begin{subeqnarray}
\frac{{\rm d}U}{{\rm d}s} = 
      I^{(a)} \left[ \frac{1}{2} (\kappa^{(1)}_{\nu}-\kappa^{(2)}_{\nu}) \sin2\chi_{B} 
      - 2 \; {\rm d}\chi_{{\rm F}}/{\rm d}s \right] \nonumber \\
     +I^{(b)} \left[ \frac{1}{2} (\kappa^{(1)}_{\nu}-\kappa^{(2)}_{\nu}) \sin2\chi_{B} 
     + 2 \; {\rm d}\chi_{{\rm F}}/{\rm d}s \right] \nonumber \\
     -\kappa_{\nu} U - (\varepsilon^{(1)}_{\nu}-\varepsilon^{(2)}_{\nu}) \sin2\chi_{B}
\setcounter{eqsubcnt}{0}
\label{us}
\end{subeqnarray}
with the average $\kappa_{\nu}=(\kappa^{(1)}_{\nu}+\kappa^{(2)}_{\nu})/2$. The
derivative ${\rm d}\chi_{{\rm F}}/{\rm d}s$ represents the change of the plane
of polarization per unit distance ${\rm d}s$ due to Faraday rotation.

A simpler formulation for the transfer of synchrotron radiation can be
obtained when neglecting the different polarizations (\cite{MHD97}). For the
total intensity, the emission and absorption coefficients can be computed
using, respectively 
\begin{eqnarray}
\varepsilon_{\nu} \propto p^{(\alpha+3)/2)}\nu^{-\alpha}
\label{semc} \\
\kappa_{\nu} \propto p^{(2\alpha+7)/2}\nu^{(\alpha+5/2)}
\label{sabc}
\end{eqnarray}
being $p$ the thermal pressure and $\alpha$ the spectral index. The total
intensity can then be integrated using (\cite{Ti91})
\begin{equation}
I=I_0 e^{-\tau_{\nu}} +
\frac{\varepsilon_{\nu}}{\kappa_{\nu}} (1-e^{-\tau_{\nu}})
\label{sint}
\end{equation}
where $\tau_{\nu}$ is the optical depth.

Further simplifications can be considered by ignoring opacity effects, in
which case an estimation of the total intensity emission can be obtained just
by adding the emission coefficient (Eq. \ref{semc}) along the different cells
in the line of sight (\cite{KF97}).

\subsection{Relativistic Effects}
\label{re}

The presence of emitting gas at velocities close to that of the speed of light
enhance the importance of the relativistic effects in the final emission
structure of the simulated maps. The emission and absorption coefficients to
be used in Eqs. (\ref{ia}-\ref{us}) are those transformed into the observer's
frame using the standard Lorentz transformations
\begin{eqnarray}
\varepsilon^{ob}_{\nu^{ob}}=\delta^2\varepsilon_{\nu} \\
\kappa^{ob}_{\nu^{ob}}=\delta^{-1}\kappa_{\nu}
\end{eqnarray}
where $\delta =\Gamma^{-1}(1-\beta \cos \theta )^{-1}=\nu^{ob}/\nu$ is the
Doppler factor; $\theta$ the viewing angle; $\beta$ the flow velocity in units
of the speed of light; and $\Gamma$ the flow bulk Lorentz factor. Note that
light aberration (see e.g., \cite{RL79}) changes the orientation of the line
of sight as seen in the fluid's frame, and therefore the relative orientation
of the magnetic field and line of sight as seen in the fluid frame,
$\vartheta$. The emission and absorption coefficients are a function of
$\sin\vartheta$ (Eqs. \ref{ecf} and \ref{acf}), and therefore light aberration
can significantly affects the synchrotron total and polarized emission as a
function of the flow velocity or viewing angle (see section \ref{jst}).

Besides light aberration, time delays is the most important effect determining
the final emission structure (no superluminal motions can be obtained from
these simulations without considering the time delays between different jet
regions). Provided the hydrodynamical variables are cell and time dependent,
to account for delays within the jet is necessary to compute the emission and
absorption coefficients at a retarded time, given by
\begin{equation}
\tau =t-{{\overrightarrow{x}.\overrightarrow{l}}\over{c}}
\end{equation}
where $\overrightarrow{x}$ is the position vector of the cell, 
$\overrightarrow{l}$ denotes the line of sight unity vector, and $c$ is the
speed of light.

We can investigate the observational consequences of light travel delays by
considering the effects produced in the emission of shocked jet material
(\cite{Al92}, \cite{JL94b}). Because of the time delays between the far and
near sides of a shock front, it appears rotated in the observed frame by an
angle $\arccos\beta$. Depending on the pattern velocity of the shock front and
viewing angle, time delays have a tendency towards aligning the shock front
with the visual. This may have relevant effects in the emission time
variability of material being heated by a shock by producing a ``phasing''
effect of the emission as measured by the observed, thus allowing for very
rapid variability (\cite{SSP98}).

Light travel delays between the forward and reverse shocks produce a
lengthening of the shocked material region in the observers frame by a factor
$\sin\theta/(1-\beta\cos\theta)$ (\cite{JL94b}, \cite{CG99}). Therefore, it is
possible to obtain estimations of the shocked material size in the source
frame from the measured sizes, velocities, and viewing angles of superluminal
components. High resolution VLBI observations (\cite{JB95}, \cite{JL98},
\cite{JL99b}, \cite{Mat98}, \cite{JLSci}, \cite{GG01}) reveal components sizes in some
cases of the order of the jet width. If we assume commonly estimated values of
$\Gamma\sim 10$ and $\theta=10^{\circ}$, this implies that the emitting
material associated with the superluminal component must be $\sim 1/9$ smaller
than the jet width. Thus, either shocks are very thin in the source frame, or
radiative losses limit the emitting region in shocks to a thin layer
(\cite{Al96b}). It is also possible that, instead, multiple superluminal
components may be associated with a single moving shock (see section
\ref{rs}).

\section{Hydrodynamical Models of Superluminal Sources}
\label{HDSS}

Shock-in-jet models (\cite{BK76}, \cite{MG85}, \cite{HAA85}) have been proven
to provide a general explanation for the emission variability observed in
components of relativistic jets. Numerical relativistic HD and emission
simulations provide a new powerful tool to improve upon these previous
analytical models. With these new numerical techniques it is now possible to
study with great detail the generation, structure, and evolution of strong
shocks, and analyze its importance in the overall dynamical evolution and
emission of jets through comparison with recent high resolution VLBI
observations.

\subsection{Relativist Shocks}
\label{rs}

Superluminal components as associated with moving shock waves have
been studied by relativistic hydrodynamical and emission models (\cite{JL97},
\cite{KF97}, \cite{MHD97}). In these models, moving shocks are induced 
by the introduction of perturbations in steady relativistic jets, studying the
subsequent jet evolution.

In \cite{JL97} the fluid jet dynamics are computed using a relativistic,
axially--symmetric jet model obtained by means of a high--resolution shock
capturing scheme (\cite{Ch95}, \cite{Ch97b}) to solve the equations of
relativistic hydrodynamics in cylindrical coordinates.  The jet material is
represented by an ideal gas of adiabatic exponent 4/3 and the quiescent state
corresponds to a diffuse ($ \rho _{b}/\rho _{a}=10^{-3} $), relativistic
($\Gamma _{b}=4 $), overpressured ($p_b\!=\!3p_a/2$), cylindrical beam with
(local) Mach number $ M_{b}=1.69 $ (subscripts $a$ and $b$ refer,
respectively, to atmosphere and beam). The jet propagates through a
pressure--decreasing atmosphere which allows the jet to expand radially. The
initial pressure mismatch in the model causes recollimation shocks and
expansions in the jet flow (\cite{JL95}). The formation and evolution of shock
waves is studied by introducing a square--wave increase of the beam flow
velocity from the quiescent value $\Gamma _{b}=4$, to $ \Gamma _{p}=10 $
during a short period of time $ \tau _{p}=0.75 R_b/c$. Because of the faster
flow velocity in the perturbation, the fluid in front piles up, creating a
shocked state, which is trailed by a rarefaction. 

\begin{figure}[t]
\begin{center}
\includegraphics[width=.95\textwidth]{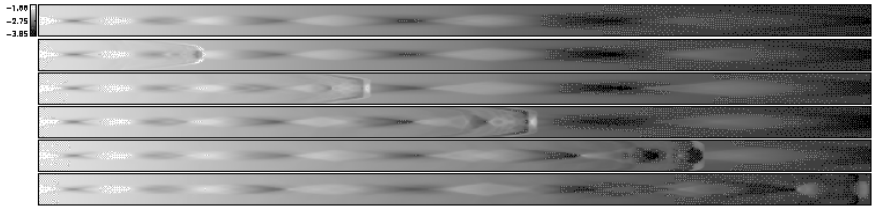}
\end{center}
\caption[]{Pressure distribution at six epochs (0 to 200 $R_b/c$
in steps of 40) after the introduction of a square-wave perturbation to the
flow Lorentz factor for the jet model discussed in the text.  The simulation
has been performed over a grid of 1600$\times$80 cells, with a spatial
resolution of 8 cells/$R_b$ in both radial and axial directions. Reprinted
from \cite{JL97}.}
\label{hdsim}
\end{figure}

The resulting dynamical evolution of the perturbation along the jet is shown
in Fig. \ref{hdsim}, which contains a set of panels showing the pressure
distribution at different epochs. The first panel corresponds to the quiescent
jet. Both the shocked and rarefied regions in the perturbation are clearly
seen. When the perturbation passes through a standing shock, the latter is
``dragged'' downstream for some distance before returning to its initial
position as the steady jet becomes reestablished.

\begin{figure}[t]
\begin{center}
\includegraphics[width=.95\textwidth]{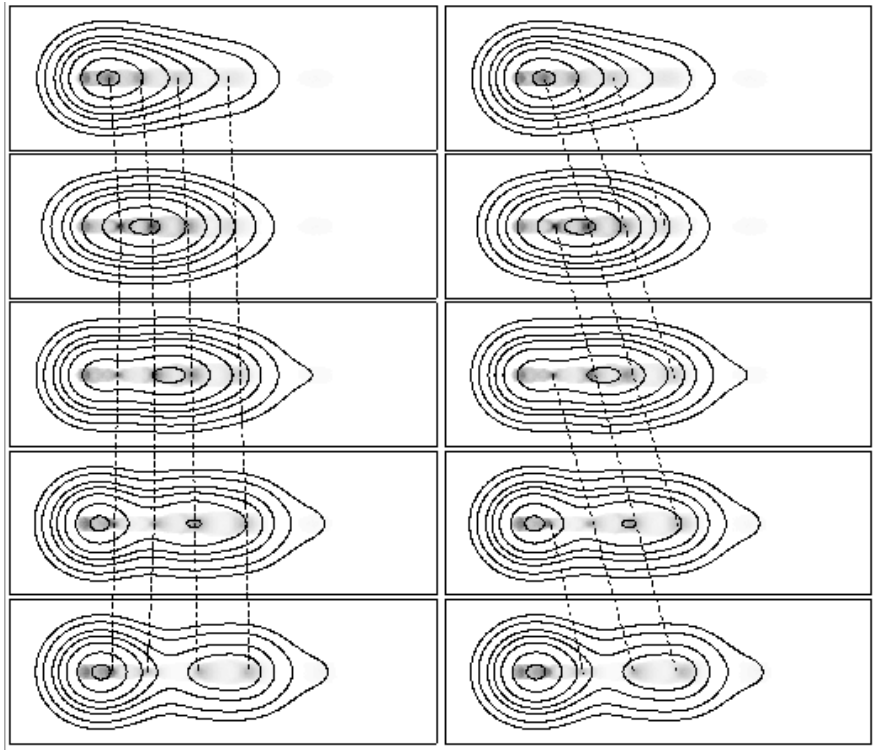}
\end{center}
\caption[]{Simulated total intensity maps of the hydrodynamical model presented
in Fig. \ref{hdsim} at five different epochs. Both, left and right image
sequences, represent the same data but with different components
identification (see text). Grey scale (normalized to the maximum of all five
epochs) shows the emission maps with the full resolution provided by the
simulations. Contours show the same images once convolved with a Gaussian beam
to resemble actual VLBI observations. Top panels show the stationary
model. Maps are obtained for an optically thin observing frequency, and a
viewing angle of $10^{\circ}$. Reprinted from \cite{JL97}.}
\label{emsim}
\end{figure}

Figure \ref{emsim} shows the total intensity maps corresponding to the
stationary model (top panels), and four epochs in the evolution of the
disturbance along the jet.  Left and right image sequences of Fig. \ref{emsim}
represent the same data, but with different components identification. By
looking at the unconvolved stationary total intensity image we observe a
regular pattern of knots of high emission, associated with the increased
specific internal energy and rest-mass density of internal oblique shocks
produced by the initial overpressure in this model. VLBI cores can be
interpreted as a first strong recollimation shock in the steady jet
(\cite{DM88}, \cite{Al98}, \cite{JL95}, \cite{Mat98}). The regular pattern of
knots should remain constant in flux and position as long as the jet inlet
hydrodynamical variables remain unchanged. Therefore, these components
resulting from the recollimation shocks may represent an alternative
explanation for the stationary jet components commonly observed in many
sources (\cite{Kr90}, \cite{We97}, \cite{JB95}, \cite{JL99c}, \cite{JL00}) as
opposed to jet bendings (\cite{Ant93}, \cite{JL99a}).

The time evolution of the convolved maps in Fig. \ref{emsim} shows the usual
core--jet VLBI structure of a blazar, with a single well--defined traveling
component associated with the moving shock. The unconvolved maps show a much
more complex jet structure. Due to time delays, the shocked region appears as
a very extended region of higher emission (see section \ref{re}), which is
moving and interacting with the quiescent jet. A tentative identification of
components through epochs is shown in the right sequence of images of
Fig. \ref{emsim}, where components are connected by dashed lines. Without
further information from the simulations, this would seem the most plausible
identification of components, since it would conclude the existence of
multiple superluminal components with similar apparent motions to that of the
main single superluminal component obtained by analyzing the lower resolution
images, that is, the convolved maps. However, this identification of
components is completely wrong. When analyzing the simulations through
intermediate epochs to those shown in Fig. \ref{emsim} we obtain the correct
identification of components, marked on the left sequence of images of
Fig. \ref{emsim}. This shows the importance of a well time sampled monitoring
when studying and identifying superluminal components through epochs. It puts
in evidence how easily a wrong identification of components may result from a
sparse time monitoring. Most of the information obtained from analyzing VLBI
images is deduced from the measured apparent motions, which, as shown here,
may easily be completely wrong, and so the obtained conclusions.

By analyzing the structural changes in the correctly identified images of
Fig. \ref{emsim} we observe that the interaction of the moving shock with the
underlying jet produces a temporary ``dragging'' of the previously stationary
features, accompanied by an increase in their fluxes. Components later on come
to a stop, followed by upstream motions of the inner components. This upstream
motion does not represent actual upstream movement of the jet fluid, but a
re-positioning of the recollimation shock closer to the jet inlet.

As the images of Fig. \ref{emsim} show, detection of this predicted dragging
and upstream motion of components requires high linear resolution
images. Some evidence of this behavior has been found in the jet of 3C~454.3
(\cite{Al98}), where 43 GHz VLBA observations have revealed the existence of a
stationary component that moves downstream slightly before returning back
upstream as a moving component passes it. Other evidence has been found in the
jets of 3C~120 (\cite{JL98}, \cite{JL99p}), 0735+178 (\cite{Ga94}), 3C~279
(\cite{We97}), and may be expected in other sources as more high--frequency
images become available.

In \cite{MHD97} the appearance of VLBI knots is studied by obtaining the total
intensity emission from relativistic flows computed using the relativistic
hydrodynamical code of \cite{DH94}. Computation of the synchrotron radiation
is obtained by computing the emission and absorption coefficients
(Eqs. \ref{semc} and \ref{sabc}), taking into account opacity effects to
integrate the transfer equation (Eq. \ref{sint}). Time delay effects are
ignored because the jet structures are found to move at barely relativistic
speeds.

\begin{figure}[t]
\begin{center}
\includegraphics[width=.95\textwidth]{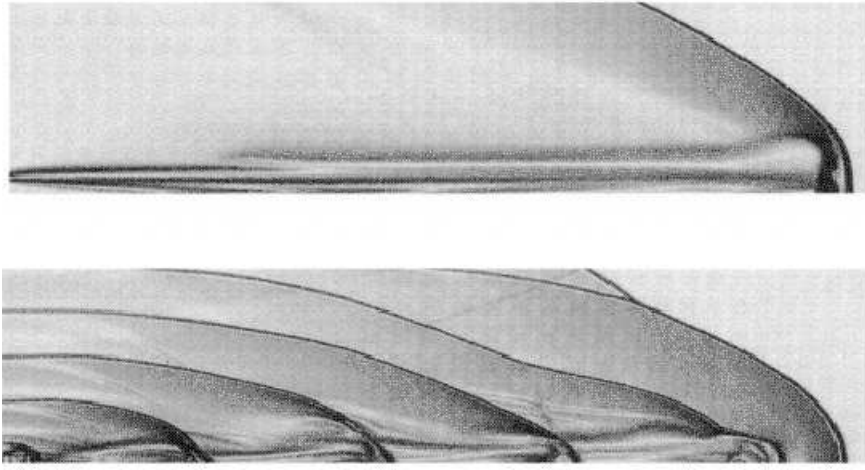}
\end{center}
\caption[]{Schlieren-type images of laboratory frame density gradient for a
jet with a Lorentz factor of 10 and adiabatic index of 4/3. Bottom image shows
the same jet after the inflow Lorentz factor has been sinusoidally modulated
between 1 and 10 to induce perturbations. Reprinted from \cite{MHD97}}.
\label{hughes}
\end{figure}

Making use of this numerical model, perturbations in the jet are studied in
\cite{MHD97} by introducing a sinusoidal modulation of the inflow Lorentz
factor between 1 and 10. Figure \ref{hughes} shows the obtained density plots
before, and after the perturbations are introduced. The relative dominance of
the intrinsic emissivity and Doppler boosting in the intensity images is
studied by computing the emission at different observing viewing angles. For
small viewing angles the image morphology is found to be determined primarily
by the Doppler boosting of the high-velocity jet, whereas at larger angles the
intrinsic emissivity is more important. Blazars are assumed to be observed
along small viewing angles, and therefore the appearance of VLBI knots is
determined primarily by the Doppler boosting of fast moving jet perturbations.

\subsection{Trailing shocks}

\begin{figure}[t]
\begin{center}
\includegraphics[width=.75\textwidth]{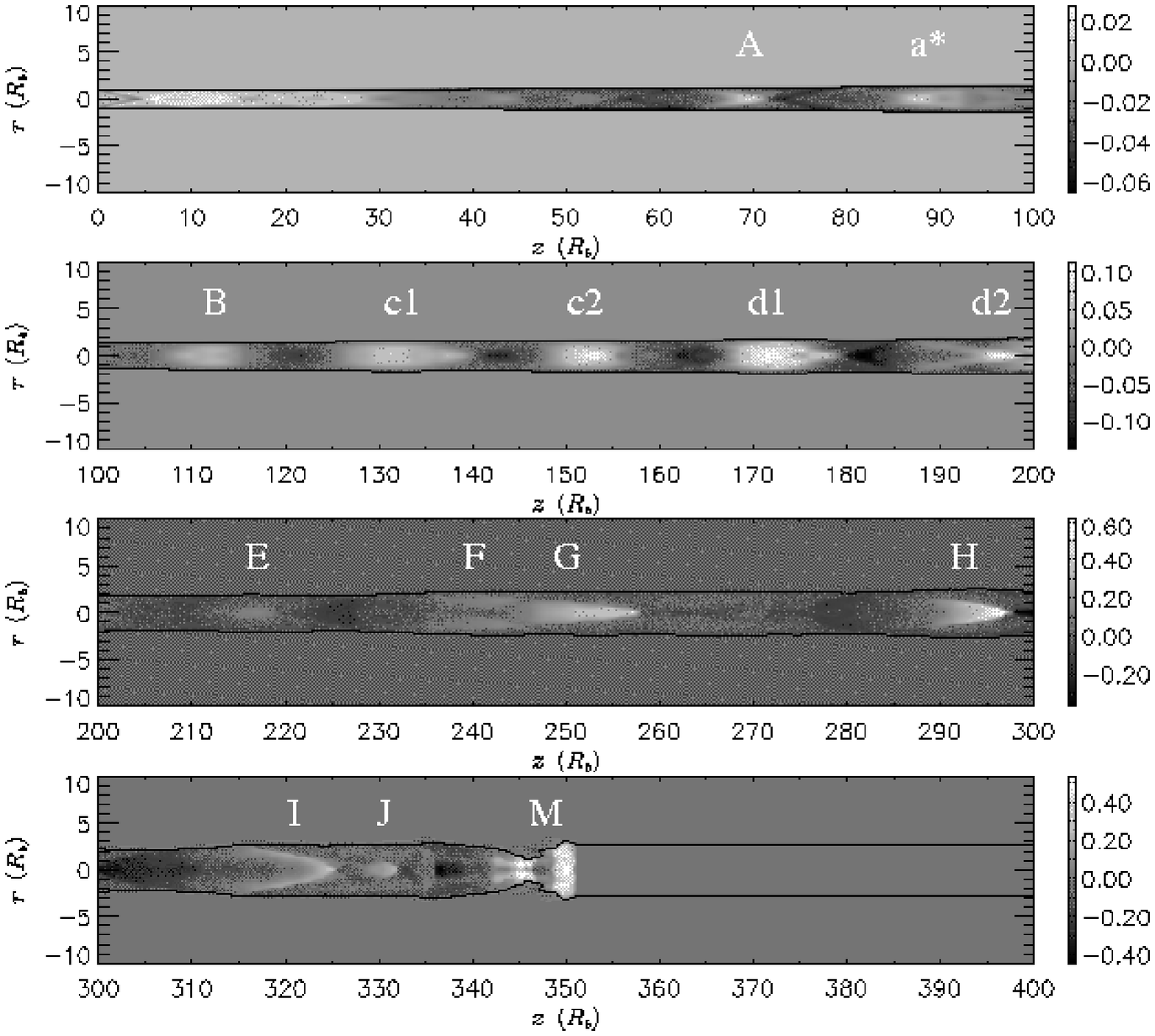}
\end{center}
\caption[]{Relative variation with respect to the quiescent jet of the Lorentz
factor (logarithmic scale). Multiple conical recollimation shocks (``trailing
shocks'') are found to follow the main shock labeled ``M''. Reprinted from
\cite{IJL01}.}
\label{trlsh}
\end{figure}

The evolution of a strong shock wave cannot ideally be isolated from the
underlying jet flow. During its motion along the jet the shock wave interacts
with the ambient jet medium, as well as the quiescent flow. This highly
non--linear interactions trigger a local pinch instability (\cite{Ha00}) that leads to
the formation of a series of conical shocks. Some of these shocks are present
in the simulations of Fig. \ref{hdsim} and have been studied in detail by
\cite{IJL01}.

Figure \ref{trlsh} shows the Lorentz factor distribution for a jet simulation
after the passage of a strong shock. Multiple conical recollimation shocks
(``trailing shocks'') can be found to follow the main perturbation. Although
their strength is a function of the distance from the jet inlet, simulated
total intensity maps show that they should be strong enough as to be
detectable by present VLBI arrays (\cite{IJL01}).

These trailing shocks can be easily distinguished because they appear in the
simulated maps as components being released on the wake of primary
superluminal component (associated with the leading shock), instead of being
ejected from the core of the jet. Those trailing components appearing closer
to the core show small apparent motions and a very slow secular decrease in
brightness, from which they could be identified as stationary
components. Those appearing farther downstream are weaker and can reach
superluminal apparent motions. Their oblique nature should also result in
different polarization properties from that of the main planar leading
shock. The existence of these trailing components indicates that not all
observed components necessarily represent a major perturbation at the jet
inlet; rather, multiple emission components can be generated by a single
disturbance in the jet.

A sample of 42 $\gamma$--ray blazars observed at high frequencies with the
VLBA has revealed that stationary components are more common than previously
thought (\cite{Sv01}). In 27 of those sources at least one non--core
stationary component has been observed. By analyzing the properties of these
stationary features two different classes of stationary components are
established (\cite{Sv01}): those within about 2 mas of the core, probably
associated with standing hydrodynamical compressions, and those farther down
the jet, probably associated with bends in the jet. These inner stationary
features are in good agreement with the properties predicted for the trailing
shocks, and therefore their association seems a plausible interpretation for
their nature. Polarimetric high resolution VLBI observations should provide
the necessary information as to confirm or rule out this hypothesis.

\subsection{Jet Instabilities and the Formation of Knots}

Relativistic jets in AGNs and microquasars are thought to be subject to
instabilities, perhaps due to changes in their feeding from unstable accretion
disks. These jet instabilities have been studied with great detail by linear
stability analysis of the linearized fluid equations and by non--linear
hydrodynamical simulations (e.g., \cite{Bo95}, \cite{Ch97b}, \cite{Ha98},
\cite{Ro99}, \cite{Ha00}, \cite{Xu00}).

Numerical simulations by \cite{Xu00} have revealed that mode--mode
interactions in 3D, such as helical surface and helical body mode interactions
and coupling to pinch modes, may lead to the formation of relatively
stationary knots along the jet beam. In particular, wave--wave interactions
are shown to lead to the formation of internal to the jet beam nearly
stationary knots close to the jet inlet, but to move and develop shock spurs
at larger distances. These mode--mode interactions, as well as the trailing
shocks, may explain some of the puzzling knots evolution observed in the
galactic superluminal GRO~J1655-40 (\cite{HR95}).

\section{Magnetic Fields in Relativistic Jets}

Although recent polarimetric VLBI observations are providing added information
on the magnetic field strength and structure at different jet scales, it is
still largely unknown the role played by the magnetic field in the jet
dynamics. In order to have a dynamically important magnetic field we should
look for jet regions where the magnetic pressure $B^2/8\pi$ dominates over the
thermal jet plasma pressure. This can be found in the inner jet regions, where
magnetic pressure should be of importance for the initial jet formation and
collimation.

\subsection{Formation, Collimation, and Acceleration of Jets}

Observation of the inner jet regions, where jets are formed, collimated and
accelerated, requires of the highest possible linear resolution in terms of
the black hole Schwarzschild radii, which determines the scale length for the
system. It is therefore in nearby sources with known massive central black
hole where high frequency VLBI observations can provide the necessary linear
resolution. This has been achieved by global 43 GHz VLBI observations of the
jet in M87 (\cite{JBL99}), revealing that the strong collimation of the jet
takes place at 30-100 Schwarzschild radii ($r_s$) from the black hole,
continuing out to $\sim$ 1000 $r_s$.

Thanks to the development of recent general relativistic magnetohydrodynamic
(GRMHD) numerical codes (\cite{KSK98}, \cite{KSK99}, \cite{Koi00}) it is now
possible to study the production of relativistic jet by numerical simulations
(see e.g., reviews by \cite{Me01}, \cite{MKU01}). The common scenario for jet
production requires a differentially rotating accretion disk surrounding the
massive central object. The disk is also threaded with an axial magnetic field
of sufficient strength to exert a braking force on the rotating plasma,
removing angular momentum and transfering it along the magnetic field
lines. These rotating magnetic twists push out and pinch the plasma into a
jet. This sweeping pinch mechanism appears to be nearly universal
(\cite{MKU01}).

\begin{figure}[t]
\begin{center}
\includegraphics[width=.85\textwidth]{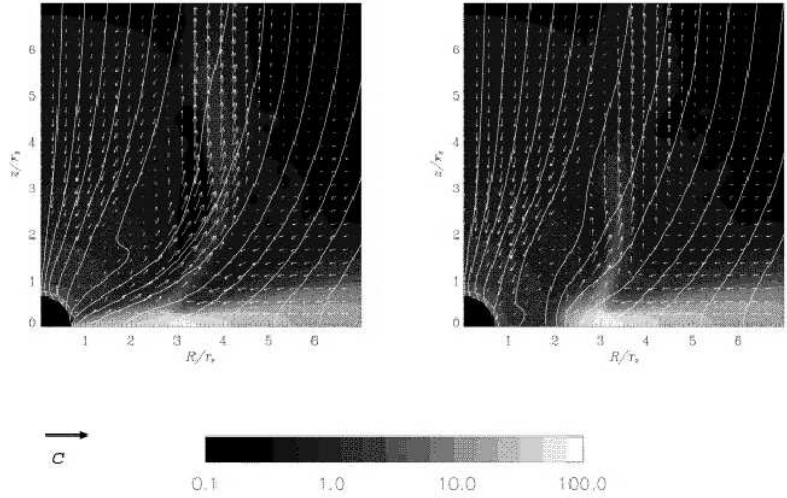}
\end{center}
\caption[]{Numerical models of jet formation for the case of a
counter--rotating (left) and co--rotating disk (right). Grey scale shows the
logarithm of the proper mass density; vectors indicate velocity; solid lines
show the poloidal magnetic field. Reprinted from \cite{Koi00}.}
\label{jform}
\end{figure}

Numerical GRMHD simulations of jet formation in a rapidly rotation Kerr black
hole have been performed for the cases of a co--rotating and counter--rotating
Keplerian accretion disk (\cite{Koi00}, Fig. \ref{jform}). For the
co--rotating disk case, a pressure driven jet is formed by a shock in the
disk, together with a weaker magnetically driven jet outside the pressure
driven jet. However, for the counter--rotating disk case, a powerful
magnetically driven jet is formed inside the pressure driven jet. This
magnetically driven jet is accelerated by a strong magnetic field created by
frame dragging in the ergosphere of the black hole.

\subsection{Intrinsic Polarimetric Differences in Jets of AGNs}

Polarimetric VLBI observations have revealed intrinsic differences in the jets
of BL Lacertae type objects and QSOs that cannot be explained solely by
differences in the viewing angle to the flow axis. First evidence for these
intrinsic differences were observed in the polarization properties of the jets
in the Pearson--Readhead sample through VLBI observations at 6 cm
(\cite{GC92}, \cite{Ca93}). These observations have shown that the magnetic
fields in BL Lac jet components are commonly perpendicular to the jet
structural axis, while for QSOs the orientation is typically aligned to the
jet axes. Recent observations (\cite{De99}, \cite{De00}, \cite{De00b}) confirm
these differences, but also provide evidence that about 30\% of the BL Lac
sources in the sample present aligned magnetic fields, similar to those found
in QSOs. This difference in the polarization properties of BL Lacs and QSOs is
interpreted by associating the observed knots with moving transverse shocks in
jets containing mainly tangled magnetic fields. Shocks would be stronger and
more commonly observed in BL Lacs, leading to the observed perpendicular
fields in the knots. On the contrary, QSOs would be required to be less
active, with weaker shocks that would never dominate in polarization.

This larger activity in BL Lacs is also supported by the University of
Michigan long--term total and polarization monitoring program
(\cite{As99}). BL Lacs are found to be more highly variable in total flux than
QSOs, and to present quasi simultaneous variations at different
frequencies. This also suggests the existence of intrinsic opacity differences
between the two classes of objects. The analysis of the polarized light curves
is indicative of the existence of propagating shocks during outbursts. The
larger variability in BL Lacs then supports the model in which shock formation
is more frequent in BL Lacs parsec-scale jets than in QSOs.

The different activity in these two classes of objects can be interpreted in
terms of jet instabilities (\cite{As99}). In \cite{Ro99} it is found that
higher jet stability can be obtained in faster and colder jet. However,
simulations by \cite{Ch97b} show that highly supersonic jets (those in which
the kinematic relativistic effects due to high Lorentz factors dominate)
present a rich internal structure, with multiple internal shocks and extended
overpressured cocoons. Both set of simulations (\cite{Ro99}, \cite{Ch97b})
agree on finding hot jets (i.e., beams with internal energies comparable to
the rest--mass energies) the most stable. Further relativistic HD simulations
of jet stability are required to explore the space parameters to determine in
which cases jets are expected to be more or less stable, and then establish a
relationship with the jets in QSOs and BL Lacs (presumably less stable).

It is also possible that this apparent differences in the jet stability of BL
Lacs and QSOs depend on the jet scales studied. Higher resolution (1.3 cm and
7 mm) polarimetric VLBI observations (\cite{Mat98}, \cite{Mat00}), therefore
exploring inner jet regions than those mapped at 6 cm, reveal no significant
differences in the polarization properties of BL Lacs and QSOs. Furthermore,
comparison between radio and optical reveals a strong correlation in the
polarization of the radio core and overall optical polarization of the source,
suggesting a common and possibly co--spatial origin for the emission at theses
frequencies. Magnetic fields perpendicular to the flow direction are commonly
observed for the radio cores. Similar orientations are found in the optical,
suggesting that the emission at both wavelengths is originated by a strong
transverse shock, perhaps the first recollimation shock in the jet, associated
with the radio core (see section \ref{rs}).

Although no significant differences in the polarization of BL Lac and QSOs are
found in the high resolution observations of \cite{Mat00}, the previous
dichotomy is translated to high-- and low--optically polarized compact
radio--loud quasars (HPQs and LPRQs, respectively). LPRQs are found to have
components with magnetic fields predominantly parallel to the jet, while in
HPQs components tend to have perpendicular magnetic field orientations. This
is interpreted assuming that LPRQs represent a quiescent phase of blazar
activity, in which the inner jet does not contain strong moving shock waves.

\subsection{Intraday Polarization Variability}

Rapid variations, with time scales less than a day, in both total and
polarized flux density have been observed in several radio sources (see
e.g. review by \cite{WW95}). If intrinsic and resulting from incoherent
synchrotron radiation, this intraday variability (IDV) implies jets with bulk
Lorentz factors between approximately 30 to 100, larger than the largest
values inferred from superluminal motions, and requiring implausibly high
brightness temperatures (\cite{BRS94}). Although most IDV at radio wavelengths
probably includes some contribution from propagation effects (\cite{Ri95}),
recent polarimetric VLBI observations reveal that some of these variations may
be intrinsic to the sources (\cite{Ga00a}, \cite{Ga00b},
\cite{Ga00c}). 

One of the first sources found to exhibit IDV is the BL Lac object
0716+714. VLA observations of this source revealed a rotation of the
polarization angle by about 50$^{\circ}$ in 12 hours. By comparison with
simulations VLBI observations, it was possible to determine that, contrary to
what it was expected, the region responsible for this variability was not the
core, but probably a feature located at about 25 milliarcseconds from it
(\cite{Ga00a}). Further IDV in polarization, but {\it not} in total flux, have
been found in the inner regions of the jets in several other sources,
including 0917+624, 0954+658, 1334-127, 2131-021, and 2155-152 (\cite{Ga00b},
\cite{Ga00c}). In the case of 2155-152, IDV variations were seen directly in the
polarized intensity images of this source at 5 GHz, where only one of the two
polarized milliarcsecond scale features varied. This represents one of the
first evidence that IDV in polarization is intrinsic to the source. Although
propagation of shocks through turbulent jets may explain some of the observed
IDV properties (\cite{Al92}), further observations and theoretical modeling
are necessary to obtain a more detailed picture of the jet physical processes
required to explain the exhibited IDV.

\section{Jet Environments}

Propagation of jets is greatly determined by the distribution of gas in their
host galaxies. As the same time, the interaction of the jet with the ambient
gas may play an important role in determining some of the observational
properties of the emission-line gas.  Distinct signs of interaction between a
collimated radio jet and a clumpy Narrow Line Region (NLR) are commonly found
in the form of morphological associations between radio and optical structures
(\cite{Ca95}, \cite{Fo98}, \cite{Be98}, \cite{Mo98}). The radio-optical
association suggests that the interaction of the jets with the interstellar
medium strongly influences the dynamics of the ionized gas in the
NLR. Furthermore, the ambient gas can be ionized by the direct interaction
with the jet bow shock, or by diffuse photoionizing radiation fields produced
in the shocks generated by such interactions, as observed in 3C277.3 and 3C171
(\cite{Ta00}).

Exploration of the time-dependent interaction of jets with the NLR have been
performed by numerical non-adiabatic hydrodynamical simulations
(\cite{Wol97a}). These simulations show that the association between the radio
and optical emission can be explained as a natural consequence of the
expansion of a hot jet cocoon into the interstellar medium. Radiative losses
create an envelope of dense cool gas and discrete emission-line knots which
can be associated with the narrow-line clouds themselves. Some of these clouds
might be partially neutral and represent sites of jet-induced star formation
(\cite{Wol97a}). Simulated $H_{\alpha}$ emission shows similar total line
widths to those observed in NLR of Seyfert galaxies, presenting large-scale
variations in the radial velocities of the clouds due to the stratified
pressure in the bow shock region of the jet (\cite{Wol97b}).

Direct collisions between the jet and clouds of the BLR and NLR are
statistically expected, depending on the assumed values for the cloud sizes
and the filling factor (e.g., \cite{Man98}). Three-dimensional numerical
hydrodynamical simulations (\cite{Wan00}, \cite{Pi99}) have been used to study
direct collisions of a jet with isolated clouds. These simulations show that,
although powerful jets would disperse the clouds, for off-center collisions
nonaxisymmetric instabilities are induced in the jet and can eventually
disrupt it. These interactions could explain some of the morphologies observed
in compact steeep-spectrum sources, such as the strongly bent geometries found
in some of these sources (\cite{Man98}).

\subsection{Jet-cloud Collisions in 3C~120}

\begin{figure}[t]
\begin{center}
\includegraphics[width=.95\textwidth]{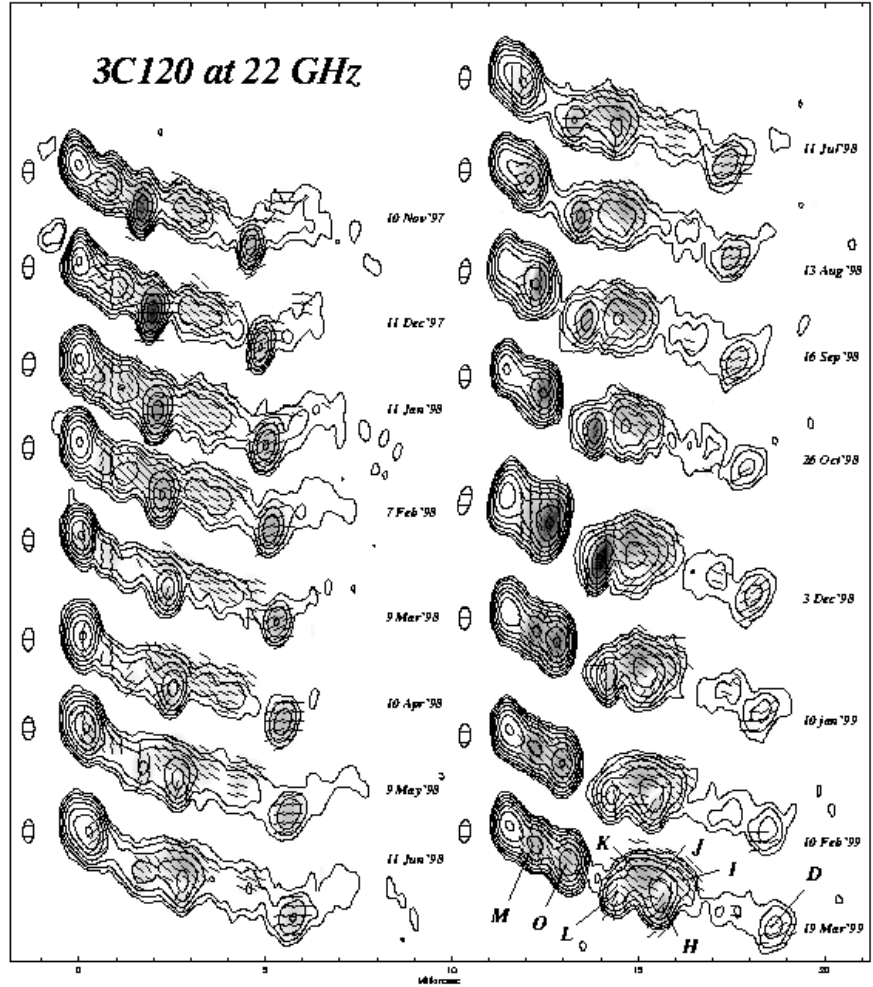}
\end{center}
\caption[]{16 epoch monthly monitoring of the jet in the radio galaxy 3C~120
obtained with the VLBA at 22 GHz. Total intensity is plotted in contours,
linearly polarized intensity in gray scale, and magnetic vectors with
bars. Reprinted from \cite{JLSci}.}
\label{3c120}
\end{figure}

The radio galaxy 3C~120 was one of the first four sources in which
superluminal motion was detected on the scale of parsecs (\cite{Se79}) to tens
of parsecs (\cite{Wa97}). High resolution polarimetric VLBI observations
(\cite{JL98}, \cite{JL99b}) have revealed a richer, more rapidly changing
structure in total and linearly polarized intensity than that found in other
relatively nearby compact extragalactic jets (\cite{JBL99},
\cite{Ti95}, \cite{Ke98}). Thanks to its proximity ($z$=0.033), millimeter VLBI
observations allow to probe the inner jet structure of 3C~120 with very high
linear resolution, $\sim$ 0.1 $h_{65}^{-1}$ pc at 43 GHz. This provides enough
resolution as to test some of the predictions obtained with the numerical
simulations outlined in section \ref{HDSS}. Towards this aim, the radio galaxy
3C~120 has been studied with unprecedented spatial and temporal resolution by
performing a 16 epoch monthly monitoring with the VLBA at 22 and 43 GHz in
dual polarization. This represents the most thorough study of a relativistic
jet to date, complete with the highest resolution and polarization
(\cite{JLSci}). The obtained sequence of images at 22 GHz is reproduced from
\cite{JLSci} in Fig. \ref{3c120}.

The images show the appearance of a new strong superluminal component, labeled
``O'' in Fig. \ref{3c120}, coincident with a major outburst in the light
curve. The passage of this new superluminal component triggered the appearance
of a stationary feature (``M'' in Fig. \ref{3c120}) that presents also
enhanced linearly polarized emission. This behavior is in agreement with the
numerical simulations of ``trailing shocks'' by \cite{IJL01}, which explain
the appearance of this stationary feature as a consequence of the jet
instabilities produced by the passage of the strong leading shock, which would
be associated with component ``O''.

\begin{figure}[t]
\begin{center}
\includegraphics[width=.85\textwidth]{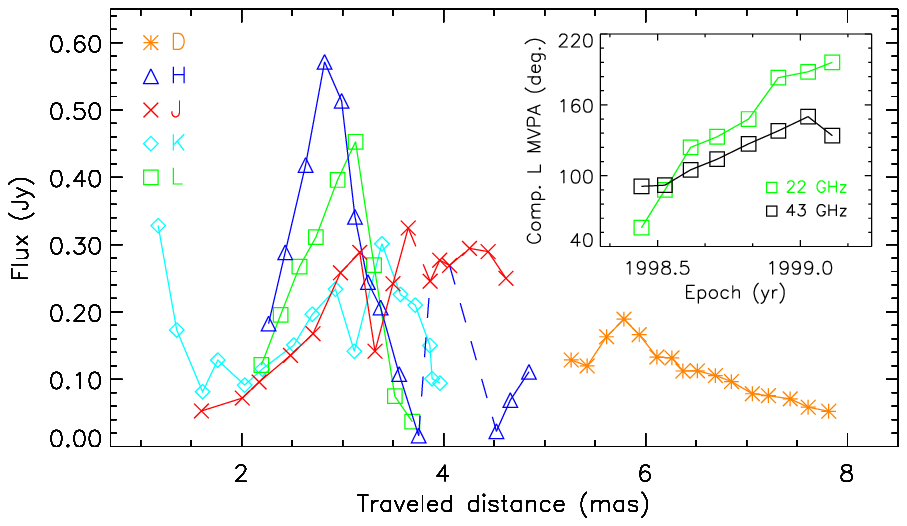}
\end{center}
\caption[]{Light curves for components in the radio galaxy 3C~120 shown in
Fig. \ref{3c120}. Inset panel shows the evolution with time of the magnetic
vector position angle of component ``L''. Reprinted from \cite{JLSci}.}
\label{3c120lc}
\end{figure}

Figure \ref{3c120lc} shows the light curves for several of the components
found in 3C~120 where a remarkable brightening can be observed starting when
components reach a distance from the core of $\sim$ 2 mas. The most pronounced
(in terms of change in flux density) flare corresponds to the component
labeled ``L'', which increased its total flux density by a factor of 9,
becoming the strongest feature in polarized intensity (Fig. \ref{3c120}). This
flare is accompanied by rotation of the magnetic vector and an increase in
degree of polarization at both, 22 and 43 GHz. The slower rotation of the
magnetic vector at 43 GHz reveals a progressive increase in the rotation
measure (RM) of component ``L'', reaching a value of $\sim 6000\pm 2400$ rad
m$^{-2}$ at peak emission.

The rapid flares in the flux densities of components are followed by equally
fast declines when they reach $\sim 3$ mas from the core. It is difficult to
explain such rapid changes in the total and polarized flux density, as well as
polarization angle, for components located between a deprojected distance of 4
and 10 pc from the core. The bending of the jet appears to be too slight to
cause such variations in brightness (from changing relativistic beaming of the
radiation relative to the observer) without an accompanying acceleration of
the apparent velocity and presence of a stationary component at this site
(\cite{Ant93}, \cite{JL94b}). Rather, interaction between the jet and a dense
cloud in the external medium seems the most plausible explanation. Similar
interactions between the jet and interstellar medium were inferred previously
from the discovery of an emission-line counterpart to the more extended radio
jet in 3C~120 (\cite{Hu88}, \cite{Ax89}). It appears that this interaction is
most intense along the southern border of the jet, where the gentle northward
curvature causes components with higher than average momentum to collide with
the external medium or cloud. Indeed, it is at the beginning of this bend that
component ``L'', which is the closest to the southern jet border and the one
exhibiting the largest flare, began to increase its flux density. This
behavior is explained if the magnetic field and population of relativistic
electrons in component ``L'' were enhanced by the shock wave produced by
interaction of the jet with the external medium, resulting in a rapid rise in
synchrotron emission. The observed increase in the degree of polarization is
then explained as a consequence of ordering of the field by the shock
wave. The rotation of the magnetic vector observed in component ``L'' can be
interpreted as Faraday rotation, the level of which can be estimated from the
different polarization angles observed at 22 and 43 GHz
(Fig. \ref{3c120lc}). After removing this effect, the relative orientation of
the magnetic field and velocity vector (which rotates as the component follows
the bend in the jet) remains at 40$\pm$10$^{\circ}$. The observed Faraday
rotation can be explained by an ionized cloud along the line of sight that may
also physically interact with the jet.

For a cloud at a temperature of 10$^4$K, free-free absorption provide an
estimated electron density of $\sim 5 \times 10^4$ cm$^{-3}$. The observed RM
of $\sim 6000\pm2400$ rad m$^{-2}$ would then require a magnetic field
strength of $\sim$ 0.4 mG. Similarly large RMs have been found in several
extragalactic jets (\cite{Ud97}, \cite{Ta98}, \cite{Ju99}), with estimated
magnetic fields of the same order. This electron density and distance from
the central engine is intermediate between those of the broad and narrow
emission--line clouds in AGNs. Given its high column density, $\sim 6\times
10^{22}$ cm$^{-2}$, such a cloud could be detected in absorption if there is a
substantial neutral atomic or molecular component, as expected. Such an
observation, which could be carried out with the VLBA in spectral-line mode,
would determine the radial velocity of the cloud and therefore whether it is
moving toward or away from the central engine.

\begin{figure}[t]
\begin{center}
\includegraphics[width=.85\textwidth]{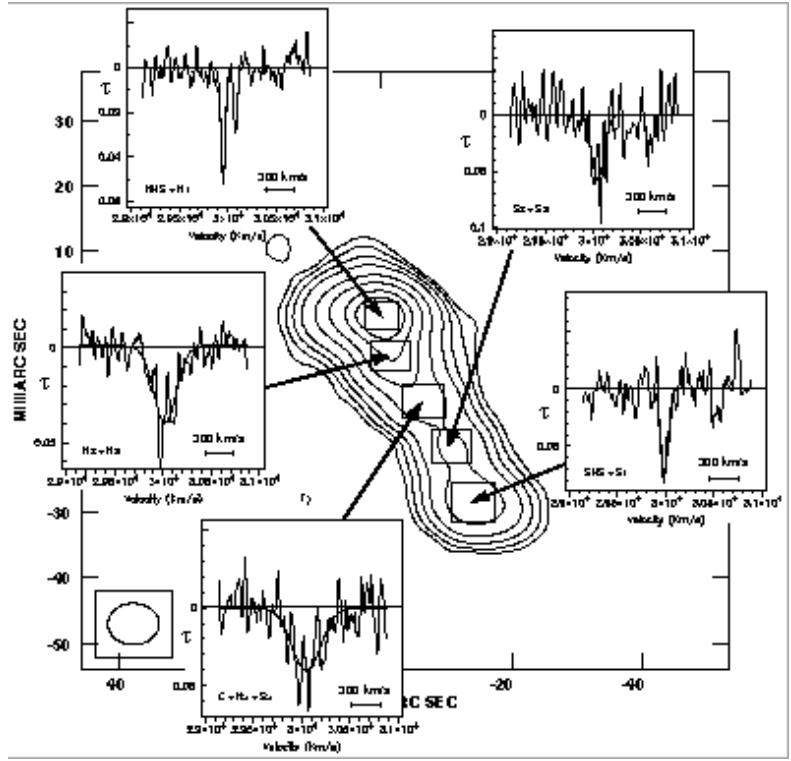}
\end{center}
\caption[]{Integrated profiles of HI absorption in five regions of the compact
symmetric object 1946+708. Contours show the radio continuum image at 1.29
GHz. Reprinted from \cite{Pe99}.}
\label{peck}
\end{figure}

Spectral--line VLBI observations have allowed to investigate the ambient
medium in AGNs with great detail (\cite{Mu95}, \cite{Ty96},
\cite{Ga98}, \cite{Pe99}, \cite{Oo00}). The common scenario outlined by these
observations consists of an accretion disk or torus surrounding the central
engine of the AGN. Depending on the source geometry, part of the jet radio
continuum would be absorbed by the atomic gas that mainly comprises the disk,
producing the HI absorption lines. The UV photons from the central engine
would ionize the inner gas of the AGNs, leading to free-free absorption of the
jet radio continuum.

In the compact symmetric object 1946+708, VLBA spectral--line observations
(\cite{Pe99}) have revealed narrow HI absorption lines in the jet northern hot
spots, indicative of small clouds of warm neutral gas associated with an
extended clumpy torus located between the radio jet and the observer (see
Fig. \ref{peck}). The high velocity dispersion and column density toward the
core of 1946+708 suggests fast moving material, possibly in rotation around
the central engine.

\subsection{Jet Stratification}
\label{jst}

Propagation of relativistic jets through the ambient medium leads to the
formation of shear layers. Such such layers have been invoked in the past to
account for a number of observational characteristics observed in FR I
(\cite{La96}, \cite{La99}) and FR II sources (\cite{Sw98}).

\begin{figure}[t]
\begin{center}
\includegraphics[width=.8\textwidth]{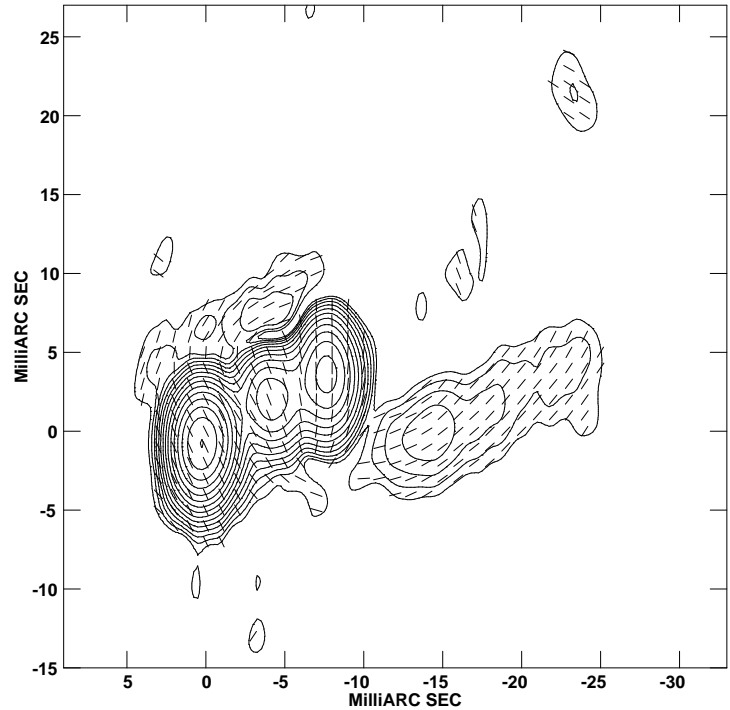}
\end{center}
\caption[]{Linear polarization distribution of the blazar 1055+018 obtained
with the VLBA+Y1 at 5 GHz. The ticks show the magnetic vector
orientation. Reprinted from \cite{ARW99}.}
\label{1055}
\end{figure}

One of the best observational evidence for these shear layers have been
recently found in the arcsec scale jet of 1055+018
(\cite{ARW99}). Polarization imaging with the VLBA at 6 cm shows that 1055+018
apparently consists of i) a emission spine along the jet axis containing a
series of knots in which the magnetic field is predominantly perpendicular to
the axis, and ii) a boundary layer in which the magnetic field is
predominantly parallel to the axis, as shown in Fig. \ref{1055}. The aligned
magnetic field in the shear layer is assumed to be originated by the jet
interaction with the ambient gas. This cross-section asymmetry presents
however a rather unusual structure, since it is observed to change with
distance along the jet: the shear layer is only visible on one side of the jet
at a time.

Three dimensional numerical hydrodynamic and emission simulations have been
performed to investigate the formation of shear layers and their implications
in the jet emission (\cite{MA00}). These simulation show that the interaction
of the jet with the external medium gives rise to a jet stratification in
which a fast spine is surrounded by a slow high-energy shear layer. In order
to explore the polarization observational properties of such a jet
stratification, \cite{MA00} considered an ad hoc distribution of the magnetic
field consisting of two components: i) a toroidal field present both in the
jet spine and the shear layer, and ii) a second component aligned in the shear
layer and radial in the jet spine. The resulting projected magnetic field is
aligned in the shear layer and is perpendicular in the jet spine, as suggested
by several observations (\cite{La96}, \cite{Sw98}, \cite{ARW99}).

\begin{figure}[t]
\begin{center}
\includegraphics[width=1.\textwidth]{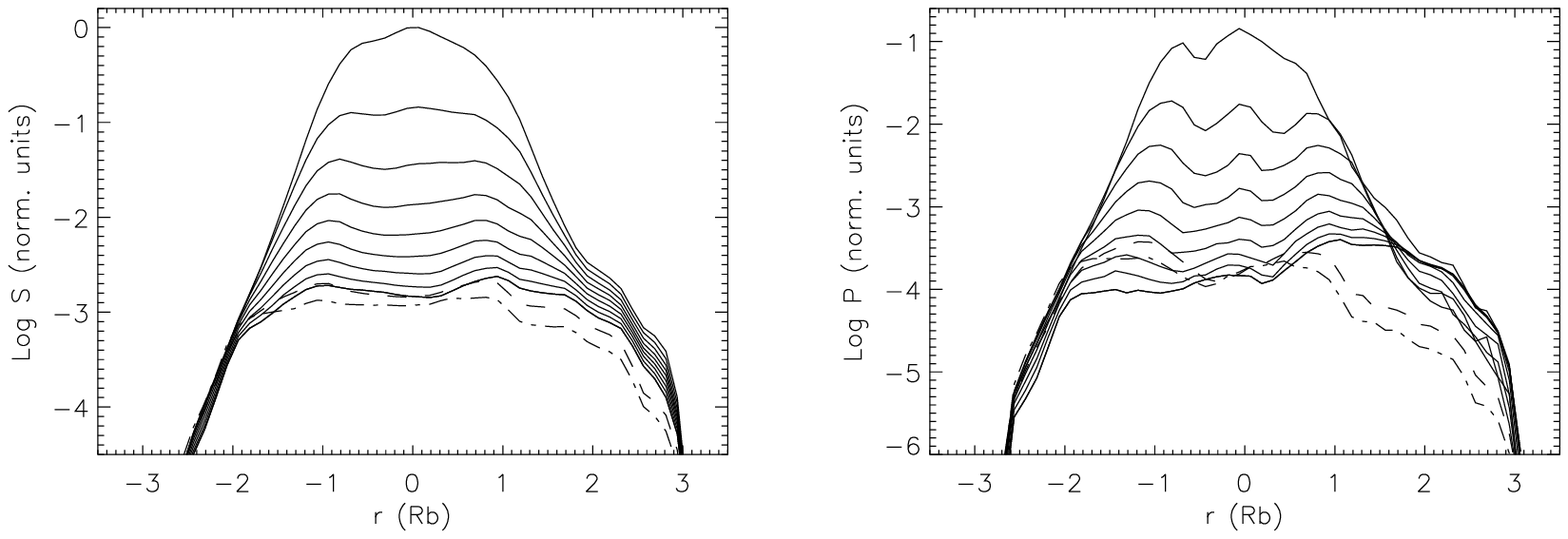}
\end{center}
\caption[]{Logarithm of the integrated total (left) and polarized (right) 
intensity across the jet for different viewing angles. Lines are plotted in
intervals of 10$^{\circ}$ between an angle of 10$^{\circ}$ (top line in both
plots) and 90$^{\circ}$ (showing a progressive decrease in emission). Dashed
(dot dashed) lines correspond to an observing angle of -130$^{\circ}$
(-170$^{\circ}$). Reprinted from \cite{MA00}.}
\label{jetas}
\end{figure}

Because of this helical magnetic field structure in the shear layer, an
asymmetry in the emission is found to appear across the jet. This asymmetry is
more pronounced in the polarized emission, and is a function of the viewing
angle, as shown in Fig. \ref{jetas}. The synchrotron radiation coefficients
are a function of the sine of the angle between the magnetic field and the
line of sight in the fluid frame, $\vartheta$ (Eqs. \ref{ecf} and
\ref{acf}). Therefore, asymmetries in the distribution of $\vartheta$ will be
translated into the emission maps, giving rise to the jet asymmetry. In order
to compute $\vartheta$ is necessary first to Lorentz transform the line of
sight from the observer's, $\theta$, to the fluid's frame $\theta'$ (see e.g.,
\cite{RL79}). For a helical magnetic field with a pitch angle $\phi$,
measured with respect to the jet axis, the angles $\vartheta^t$ and
$\vartheta^b$ (where superscripts $t$ and $b$ refer to the top and bottom of
the jet, respectively) add $2 \phi$ (note that $\vartheta^{t,b}$ is always
defined as positive). Therefore, as long as $\phi$ is different from zero or
$\pi/2$, i.e. the field is neither purely aligned nor toroidal, the factor
$\sin\vartheta^{t,b}$ in the synchrotron radiation coefficients will introduce
an asymmetry in the jet emission. This asymmetry will reach a maximum value
for a helical magnetic field with $\phi=\pi/4$. However, indistinctly of the
helix pitch angle, the predominance between $\sin\vartheta^t$ and
$\sin\vartheta^b$ will reverse at $\theta'=\pi/2$, which corresponds to a
viewing angle in the observer's frame of $\cos\theta_r=\beta$. For a helical
field oriented clockwise as seen in the direction of flow motion (i.e., the
aligned component of the field is parallel to the jet flow), and for
$\theta'<\pi/2$ the bottom of the jet will show larger emission, while for
$\theta'>\pi/2$ the top of the jet will be brighter (the opposite is true for
a helical field oriented counter-clockwise, i.e. $\phi > \pi/2$). The maximum
asymmetry will be obtained for $\theta'=\phi$ and $\theta'=\pi-\phi$, and the
fastest transition (with changing $\theta'$) between top/bottom emission
predominance will be obtained for $\phi$ close to $\pi/2$, i.e. when little
aligned field is present.

It is interesting to note that for $\theta \sim \theta_r$, small changes in
the jet velocity or the viewing angle will produce a flip in the top/bottom
jet total and polarized emission dominance. This model has been used by
\cite{MA00} to interpret the shear layer structure observed in 1055+018 
(\cite{ARW99}). For this, 1055+018 is required to be oriented close to
$\theta_r$ and to contain a shear layer with a helical field. In this case,
the flip in the top/bottom orientation of the polarization asymmetry in
1055+018 is interpreted as due to a jet decelation, as observed for several
components in this source (\cite{ARW99}).

\section{Conclusions}

Numerical relativistic (magneto)hydrodynamic and emission simulations have
proven to be a powerful tool to understand the physics of jets in AGNs and
microquasars through direct comparison with observations. These models are
capable of study the jet dynamics with unprecedented detail, and under similar
conditions to those in actual sources (relativistic internal energies and bulk
flow velocities). Computation of the non--thermal radio emission allows to
study the relationship between radio knots and internal shock waves. These
simulations show that the evolution of moving shock waves is greatly
determined by its interaction with other standing shock waves, as well as the
underlying jet flow and external medium. ``Upstream'' knot motions,
``dragging'' of previously stationary components, and formation of multiple
``trailing'' components after the passage of a main strong shock are some of
the predictions obtained from these simulations. First observational evidence
of these features are being obtained thanks to the recent millimeter
polarimetric VLBI observations.

The dynamical and emission evolution of jet components may be severely
affected by interactions with the external medium. An extensive monitoring of
the radio galaxy 3C~120 with the highest resolution and in polarization has
provided direct imaging of the interaction between jet components and the
external medium, resulting in rapid changes in the total and linearly polarized
emission of components. These interactions between the jet and ambient
medium may also result in a jet stratification, in which a fast spine is
surrounded by a slow high--energy shear layer, leading to an emission cross
section jet asymmetry.

Further numerical simulations, and its comparison with high resolution
observations should provide new insights towards the understanding of the
physical processes taking place in the jets of AGNs and microquasars.

\end{document}